# MgGa$_2$O$_4$ spinel barrier for magnetic tunnel junctions: coherent tunneling and low barrier height

Hiroaki Sukegawa,[1,*] Yushi Kato,[2] Mohamed Belmoubarik,[1] P.-H. Cheng,[1,3] Tadaomi Daibou,[2] Naoharu Shimomura,[2] Yuuzo Kamiguchi,[2] Junichi Ito,[2] Hiroaki Yoda,[2] Tadakatsu Ohkubo,[1] Seiji Mitani,[1,3] Kazuhiro Hono[1,3]

[1]*Research Center for Magnetic and Spintronic Materials, National Institute for Materials Science (NIMS), 1-2-1 Sengen, Tsukuba 305-0047, Japan*

[2]*Corporate Research & Development Center, Toshiba Corporation 1, Komukai-Toshiba-Cho, Saiwai-ku, Kawasaki 212-8582, Japan*

[3] *Graduate School of Pure and Applies Sciences, University of Tsukuba, Tsukuba 305-8577, Japan*

**Abstract**

Epitaxial Fe/magnesium gallium spinel oxide (MgGa$_2$O$_4$)/Fe(001) magnetic tunnel junctions (MTJs) were fabricated by magnetron sputtering. Tunnel magnetoresistance (TMR) ratio up to 121% at room temperature (196% at 4 K) was observed, suggesting a TMR enhancement by the coherent tunneling effect in the MgGa$_2$O$_4$ barrier. The MgGa$_2$O$_4$ layer had a spinel structure and it showed good lattice matching with the Fe layers owing to slight tetragonal lattice distortion of MgGa$_2$O$_4$. Barrier thickness dependence of the tunneling resistance and current-voltage characteristics revealed that the barrier height of the MgGa$_2$O$_4$ barrier is much lower than that in an MgAl$_2$O$_4$ barrier. This study demonstrates the potential of Ga-based spinel oxides for MTJ barriers having a large TMR ratio at a low resistance area product.

* Electronic mail: sukegawa.hiroaki@nims.go.jp

Magnetic tunnel junctions (MTJs) have played a central role in spintronic devices such as read heads of hard disk drives and non-volatile magnetoresistive random access memories (MRAMs) for the last two decades, and many efforts have been made to improve their performance.[1] One of the most prominent achievements which accelerated the practical applications was the realization of giant tunnel magnetoresistance (TMR) ratios by using rock-salt type MgO crystalline barrier.[2-5] The giant TMR effect is attributed to the spin-dependent coherent tunneling through the $\Delta_1$ Bloch state in MgO(001). Toward ultra-high density spin transfer torque (STT)-MRAM applications, MTJs with low resistance area ($RA$) product of around 1 $\Omega \cdot \mu m^2$ are needed as well as high TMR ratios. In MgO-based MTJs, the thickness of the MgO barrier must be reduced to a few monoatomic layers for achieving such a low $RA$, which causes substantial reduction of TMR ratios. This means the necessity of alternative barriers having a low barrier height. Doping other elements into MgO has been known to reduce the MgO barrier height; e.g., Zn [6,7] and Ti [8] dopings were reported to provide lower $RA$ values although such dopings are likely to cause reduction of TMR ratios.

Spinel oxide $MgAl_2O_4$(001) barrier also exhibits the coherent tunneling effect,[9,10] and over 300% TMR ratios at room temperature (RT) were reported in $MgAl_2O_4$-based MTJs.[11,12] Very recently, a TMR ratio at RT of 92% in an MgO/spinel-type $\gamma$-$Ga_2O_3$(001) bilayer barrier MTJ [13] and that of 120% in a Li-Mg-Al-O quaternary spinel-based barrier MTJ were reported,[14] showing the capability of the spinel based barriers for MTJs. The lattice spacing of $MgAl_2O_4$ is 4% smaller than that of MgO. In addition, the lattice spacing can further be tuned by the Mg/Al composition, leading to the highly lattice-matched interfaces with various ferromagnetic materials.[9,12] On the other hand, the experimental band gap of $MgAl_2O_4$ is reported to be 7.8 eV,[15] which is similar to that of MgO (7.58–7.8 eV);[15] therefore, no reduction in $RA$ values using an $MgAl_2O_4$ barrier was observed.[16] Some of spinel oxides other than $MgAl_2O_4$, such as $ZnAl_2O_4$, $SiMg_2O_4$ and $SiZn_2O_4$, were predicted to exhibit the similar coherent tunneling effect.[17] It also was suggested that one can tune the band gap by carefully selecting constituent cations of spinel oxides.

$MgGa_2O_4$ is known as a wide bandgap transparent semiconducting oxide.[18] The lattice constant of $MgGa_2O_4$ is 0.8280 nm,[19] which is slightly lower than that of MgO with a doubled unit cell (0.842 nm). Therefore, a good lattice matching with FeCo is expected when all the layers are grown with the (001) orientation. Importantly, the reported band gap is 4.9 eV,[18] which is lower than that of $MgAl_2O_4$; consequently, lowering $RA$ values using an $MgGa_2O_4$ barrier is expected. In this letter, we report the growth of lattice-matched Fe/$MgGa_2O_4$/Fe(001) MTJs using sputtering method. We confirmed that the $MgGa_2O_4$ layer was epitaxiallly grown with a spinel structure. A TMR ratio reached 121% at RT (196% at 4 K), indicating the coherent tunneling effect in the $MgGa_2O_4$(001) barrier. Moreover, $RA$ values as a function of the barrier thickness and current-voltage characteristics revealed a low barrier height of the $MgGa_2O_4$ barrier. Therefore, Ga-based spinels are promising as an alternative MTJ barrier having a low *RA* value and a high TMR ratio suitable for future spintronic applications.

MTJ stacks were deposited by ultra-high vacuum magnetron sputtering on an MgO single crystalline substrate using the method reported for the Fe/MgAl$_2$O$_4$/Fe(001) MTJs.[20] The typical stack structure is MgO(001) substrate/Cr (40)/Fe (100)/Mg ($t_{Mg}$ = 0 or 0.6)/MgGa$_2$O$_4$ ($t_{MgGa2O4}$)/Fe (7)/Ir$_{20}$Mn$_{80}$ (12)/Ru (10) (unit: nm). For the MgGa$_2$O$_4$ layer deposition, RF sputtering of a sintered MgGa$_2$O$_4$ target was used. Other metallic layers were deposited by DC sputtering. To obtain a flat film with a high crystallinity, each layer was *in-situ* post-annealed after the deposition at RT. For the MgGa$_2$O$_4$ layer, post-annealing temperature of 500°C was used. The Mg layer was inserted to prevent oxidation of the bottom Fe surface during the MgGa$_2$O$_4$ growth and post-annealing. After the deposition, the stacks were *ex-situ* annealed at 175°C under a magnetic field of 5 kOe along the Fe[100] direction. The stacks were then patterned into micrometer scale junctions using electron-beam lithography, photolithography, and Ar ion etching. The patterned MTJs were characterized by DC 4 probe method at RT and 4 K. The microstructure of the whole stack was evaluated by high-angle annular dark field scanning transmission electron microscopy (HAADF-STEM) using an FEI Titan G2 80-200 with a probe aberration corrector. The band gap of a sputtered 30-nm-thick MgGa$_2$O$_4$ film grown on an MgO(001) substrate was determined to be 4.7±0.1 eV by reflection electron energy loss spectroscopy, which was nearly equivalent to the bulk value of 4.9 eV.[18] Rutherford backscattering spectrometry (RBS) analysis revealed that the composition of an MgGa$_2$O$_4$ thin film was nearly stoichiometric (Mg$_{16}$Ga$_{29}$O$_{55}$).

Figure 1 (a) shows the cross-sectional HAADF-STEM image of an Fe/MgGa$_2$O$_4$ ($t_{MgGa2O4}$ = 2 nm)/Fe structure along the Fe[110] direction. The image indicates epitaxial growth from the bottom-Fe to the top-Fe layer with (001) orientation and flat MgGa$_2$O$_4$ interfaces. No in-plane misfit dislocations were found in the observed area. By averaging in-plane lattices (50 planes) and out-of-plane lattices (total 25 planes) using scale-calibrated HAADF-STEM images, the MgGa$_2$O$_4$ lattice was found to be tetragonally distorted; the in-plane (out-of-plane) lattice spacing was determined to be 0.204 nm (0.212 nm). The in-plane spacing of MgGa$_2$O$_4$ was equivalent to that of Fe, leading to the lattice-matched interfaces.

Additionally, a nano-beam electron diffraction (NBD) image taken from the area near the MgGa$_2$O$_4$ barrier is shown in Fig. 1 (b). The NBD spots from MgGa$_2$O$_4$ and Fe reveal the epitaxial relationship of Fe(001)[110]||MgGa$_2$O$_4$(001)[100]. The {022} spots from MgGa$_2$O$_4$ indicate the cation-order in the MgGa$_2$O$_4$ lattice. This means that the MgGa$_2$O$_4$ barrier has a spinel structure with tetragonal distortion although the accurate atomic configuration cannot be determined by analyzing only the present NBD pattern. The lattice constants of the MgGa$_2$O$_4$ were calculated to be 0.816 nm (in-plane) and 0.848 nm (out-of-plane). The ground-state structure of MgGa$_2$O$_4$ with an inverse spinel structure was predicted to be a tetragonal structure belonging to the space group $P4_3 22$ (or $P4_1 22$);[21]

however, the tetragonality in the MgGa$_2$O$_4$ layer of this work may be mainly introduced by compressive stress from thick Fe layers under and above the MgGa$_2$O$_4$ layer.

In the case of the MgAl$_2$O$_4$ barrier with the same post-annealing temperature (500°C), the cation-order was not observed, and thus the MgAl$_2$O$_4$ had a halved lattice constant (~0.404 nm) of the normal spinel. This suggests that the cation-order occurs in MgGa$_2$O$_4$ easier than in MgAl$_2$O$_4$ although the difference in the melting point between MgGa$_2$O$_4$ (1930°C) and MgAl$_2$O$_4$ (2122°C) is small. The tendency of the order-disorder of cation sites can be determined by the constituent elements. In fact, it is reported that behavior of the cation disordering of spinel oxides by ion radiation strongly depends on the constituent elements.[22,23]

Elemental maps of Mg, Ga, O and Fe in the MTJ using the energy-dispersive X-ray spectroscopy (EDS) are shown in Figs. 1 (c)-(g), which indicate that the barrier layer is chemically homogeneous. Neither segregation nor interdiffusion of each element is observed near the Fe/MgGa$_2$O$_4$ interfaces. Therefore, chemically sharp bottom-Fe/MgGa$_2$O$_4$ and MgGa$_2$O$_4$/top-Fe interfaces were confirmed. The Mg/Ga atomic ratio were estimated to be ~0.5, which is close to the ratio by the RBS analysis for the 30-nm-thick film (= 0.55).

Figure 2 shows the TMR ratio as a function of magnetic field for an Fe/Mg ($t_{Mg}$ = 0.6 nm)/MgGa$_2$O$_4$ ($t_{MgGa2O4}$ = 3.0 nm)/Fe MTJ measured at RT and 4 K (bias voltage ~5 mV). TMR ratio of 121% was observed at RT (196% at 4 K) in the MTJ, which are much higher than that expected from the Jullière model [24] (TMR ratio ~ 40–50% by assuming an Fe spin polarization of 0.40–0.45 [25]). Therefore, we conclude that the tunneling spin polarization is significantly enhanced by the $MgGa_2O_4$ barrier as observed in Fe/MgO/Fe(001) [5] and Fe/$MgAl_2O_4$/Fe(001) MTJs.[9,11] This is attributed to the coherent tunneling effect in the $MgGa_2O_4$(001) barrier as predicted in MTJs with spinel oxides.[17] The observed TMR ratios are comparable to those in the Fe/cation-ordered $MgAl_2O_4$/Fe(001) MTJ (117% at RT and 165% at 15 K),[9] whereas they are much lower than those in Fe/MgO/Fe(001) (180–200% at RT)[5,26] and Fe/cation-disordered $MgAl_2O_4$/Fe (~245% at RT).[20] This could be attributed to the band-folding effect of the Fe $\Delta_1$ bands due to the doubled lattice size of spinel $MgGa_2O_4$.[10] Therefore, the introduction of a cation-disorder into the $MgGa_2O_4$ barrier or the use of half-metallic ferromagnetic electrodes will be necessary to achieve a much larger TMR ratio in $MgGa_2O_4$ based MTJs, as examined in the $MgAl_2O_4$ based ones.[11,12]

Lower $RA$ values in the $MgGa_2O_4$ barrier are expected due to its low barrier height as a consequence of its lower band gap (4.9 eV)[18] than that of $MgAl_2O_4$ (7.8 eV)[15] and that of MgO (7.58–7.8 eV).[15] To evaluate the barrier heights, we plotted ln($RA$) of the parallel (P) magnetic configuration at RT as a function of the barrier thickness ($t_{barrier}$) for Fe/$MgGa_2O_4$/Fe(001) ($t_{Mg} = 0$ nm) and Fe/$MgAl_2O_4$/Fe(001) MTJs in Fig. 3. To obtain comparable MTJ stacks, we deposited wedge-shaped barriers using direct sputtering and a linear motion shutter for both the barriers. The post-annealing temperature of the barriers was fixed at 500°C. The ln($RA$) for both the cases linearly increases with $t_{barrier}$, indicating that direct tunneling is the dominant electron transport mechanism for both the MTJs. As expected, the absolute $RA$ values and the $\ln(RA)$ slope of the MgGa$_2$O$_4$ MTJs are much smaller than those of the MgAl$_2$O$_4$ MTJs. As a consequence of the smaller $\ln(RA)$ slope of the MgGa$_2$O$_4$ MTJs, the $RA$ of the MgGa$_2$O$_4$ barrier increases by a factor of 10 with a thickness increment of 0.38 nm, which is much lower than 0.49 nm of the MgAl$_2$O$_4$ barrier, as indicated in Fig. 3. These results suggest that MgGa$_2$O$_4$ has a lower barrier height than MgAl$_2$O$_4$.

To further examine the barrier height of the MgGa$_2$O$_4$ barrier, the current density ($J$)-bias voltage ($V$) curves for MTJs with the same barrier thickness were evaluated. Figure 4 (a) shows that the $J$-$V$ characteristics for parallel (P) and antiparallel (AP) configurations at RT for Fe/MgGa$_2$O$_4$ (2.4 nm)/Fe ($t_{Mg}$ = 0 nm) and Fe/MgAl$_2$O$_4$ (2.4 nm)/Fe MTJs. Here, the positive bias corresponds to electron tunneling from the bottom electrode to top one. For both the MTJs, the nearly symmetric curves with respect to the bias direction were observed. Since the $RA$ value of the MgGa$_2$O$_4$ barrier is much lower than those of the MgAl$_2$O$_4$ one (see Table 1), larger $J$ can be applied to the MgGa$_2$O$_4$-MTJs. In Fig. 4 (b), we replotted the curves for the positive bias with log scales (i.e., ln $J$-ln $V$). In general, the bias voltage where rapid $J$ increase is observed roughly corresponds to the barrier height (unit in eV). Therefore, the MgGa$_2$O$_4$ barrier evidently has a lower barrier height than the MgAl$_2$O$_4$ barrier consistently with the $RA$ values as discussed earlier. To compare their barrier heights, we also plotted the fitting results in Fig. 4 (b) using the Simmons' equation:[27]

$$J = (6.2\times10^8/t_{eff}^2)\{(\phi_{eff} - eV/2)\exp[-10.25t_{eff}(\phi_{eff} - eV/2)^{1/2}]$$
$$-(\phi_{eff} + eV/2)\exp[-10.25t_{eff}(\phi_{eff} + eV/2)^{1/2}]\}, \qquad (1)$$

where $J$ is in A/cm$^2$, $t_{eff}$ is the effective barrier thickness in nm, and $\phi_{eff}$ is the effective barrier height in eV, respectively. Note that in Eq. (1), we did not take into account the effects of asymmetry in the $J$-$V$ curves [7] and the image force [27] for simplicity. The TMR ratio, $RA$, and the fitting results were summarized in Table 1. The experimental curves were well fitted by Eq. (1), and both the MTJs are found to have similar $t_{eff}$ values (0.9–1.1 nm) in the P and AP magnetic configurations. On the other hand, the $\phi_{eff}$ values for the MgGa$_2$O$_4$-MTJ (1.0 eV for P and 1.3 eV for AP) were much lower than that for the MgAl$_2$O$_4$-MTJ (2.1 eV for P and 3.0 eV for AP). Consequently, the Simmons' fit results of the $J$-$V$ curves also confirm that the low barrier height of the MgGa$_2$O$_4$ barrier.

In summary, we demonstrated the growth of lattice-matched epitaxial Fe/MgGa$_2$O$_4$/Fe(001) MTJs. The MgGa$_2$O$_4$ barrier has a spinel structure and a tetragonal distortion was observed. The observed TMR ratio up to 121% at RT (196% at 4 K) indicates that the MgGa$_2$O$_4$(001) barrier shows the coherent tunneling effect. The barrier thickness dependence of $RA$ value and the Simmons' fit of $J$-$V$ curves suggested a low barrier height of the MgGa$_2$O$_4$ barrier compared with that of the MgAl$_2$O$_4$ barrier. Therefore, MgGa$_2$O$_4$ spinel is a promising MTJ barrier exhibiting a low $RA$, which is necessary for future high-density MRAM applications. This study shows that spinel oxides give the possibility to tune not only in their lattice constants but also in the in their barrier heights as coherent barriers of MTJs.

This work was partly supported by the ImPACT Program of Council for Science, Technology and Innovation, Japan. P.C. acknowledges National Institute for Materials Science for the provision of a NIMS Junior Research Assistantship.

**Figure captions**

FIG. 1. (a) HAADF-STEM image of an Fe/MgGa$_2$O$_4$ (~2 nm)/Fe multilayer. (b) NBD pattern taken around the MgGa$_2$O$_4$ barrier. (c)-(g) EDS elemental maps for (c) STEM image, (d) Mg, (e) Ga, (f) O and (g) Fe.

FIG. 2. TMR curves as a function of magnetic field for an Fe/Mg (0.6 nm)/MgGa$_2$O$_4$ (3.0 nm)/Fe MTJ at 4 K and 297 K (RT).

FIG. 3. ln($RA$) for the P state plotted as a function of barrier thicknesses ($t_{barrier}$) for Fe/MgGa$_2$O$_4$/Fe(001) (area: 128 μm$^2$) and Fe/MgAl$_2$O$_4$/Fe(001) MTJs (area: 39 μm$^2$) measured at RT under a bias voltage below 10 mV. Solid and dotted lines indicate the fitting results using the linear equation: ln($RA$) = $a$ + $b$ × $t_{barrier}$, where $a$ and $b$ are fitting parameters. Note that the deviation from the linear fit at the low $t_{barrier}$ region for MgGa$_2$O$_4$ is mainly due to the effect of a non-negligible electrode resistance.

FIG. 4. $J$-$V$ curves of Fe/MgGa$_2$O$_4$ or MgAl$_2$O$_4$ (2.4 nm)/Fe at RT. (a) $J$ vs. $V$, and (b) ln $J$ vs. ln $V$. In (b), dotted lines show the fitting curves using Eq. (1).

**Table**

TABLE I. Summary of experimental and fitting results of Fe/barrier/Fe MTJs with 2.4-nm-thick MgGa$_2$O$_4$ and MgAl$_2$O$_4$ barriers. $t_{eff}$ and $\phi_{eff}$ are fitting results using Eq. (1).

| Barrier/Structure | TMR ratio (%) (at $V=0$) | Configuration | $RA$ (Ω·μm$^2$) (at $V=0$) | $t_{eff}$ (nm) | $\phi_{eff}$ (eV) |
|---|---|---|---|---|---|
| MgGa$_2$O$_4$ 2.4 nm /Spinel | 80 | P | 3.02×10$^3$ | 0.96 | 1.3 |
| | | AP | 5.44×10$^3$ | 1.1 | 1.0 |
| MgAl$_2$O$_4$ 2.4 nm /Cation-disorder spinel | 236 | P | 1.37×10$^5$ | 0.89 | 3.0 |
| | | AP | 4.61×10$^5$ | 1.1 | 2.1 |

**Figures**

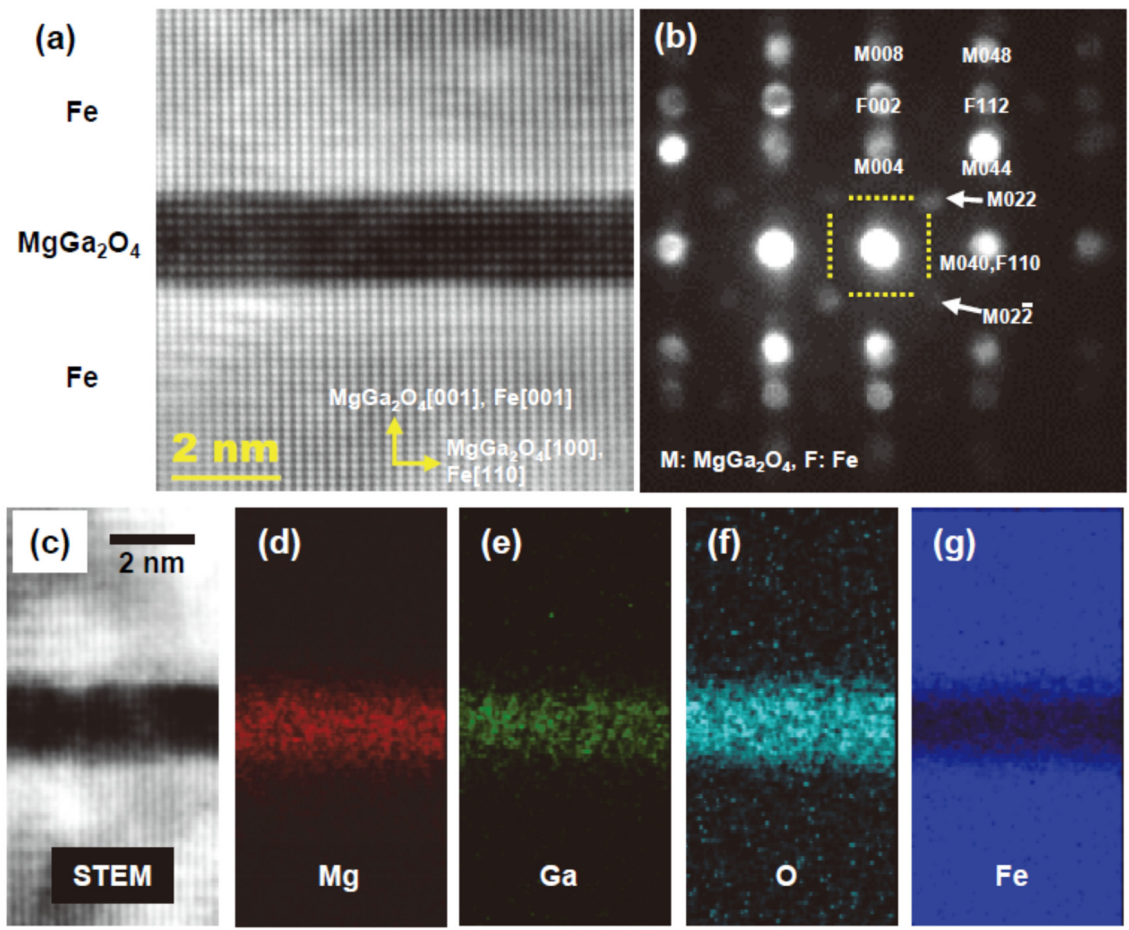

FIG. 1.

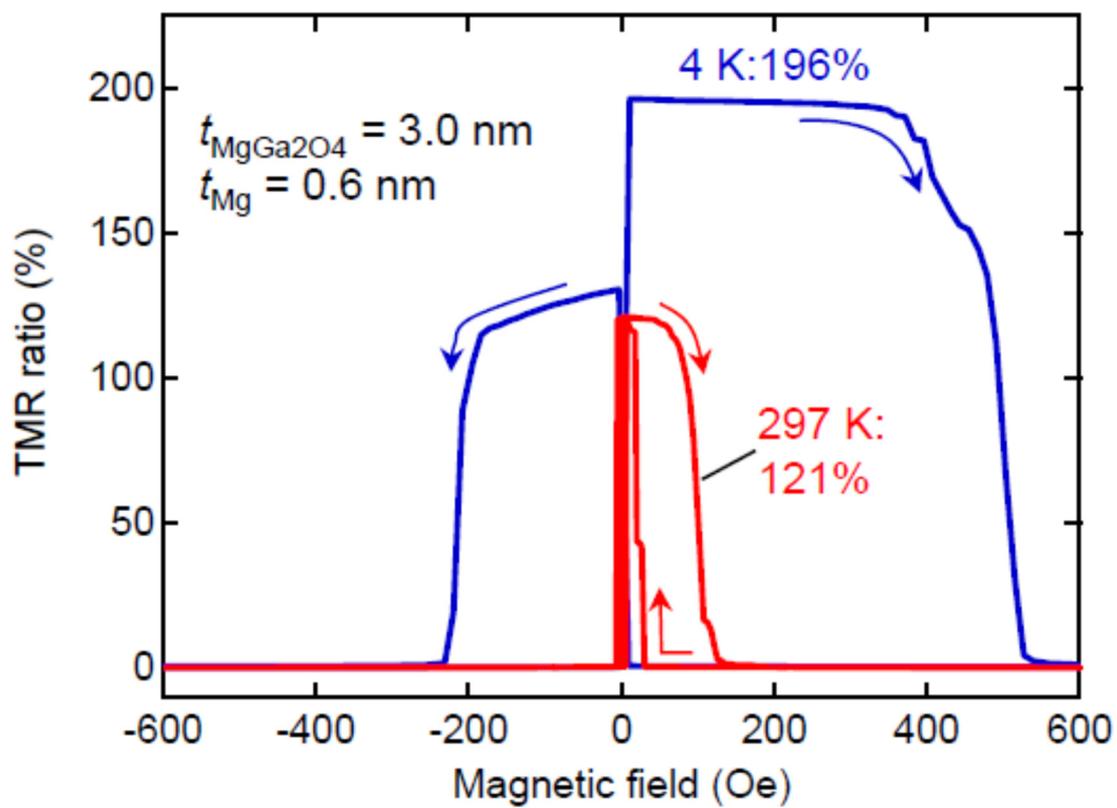

FIG. 2.

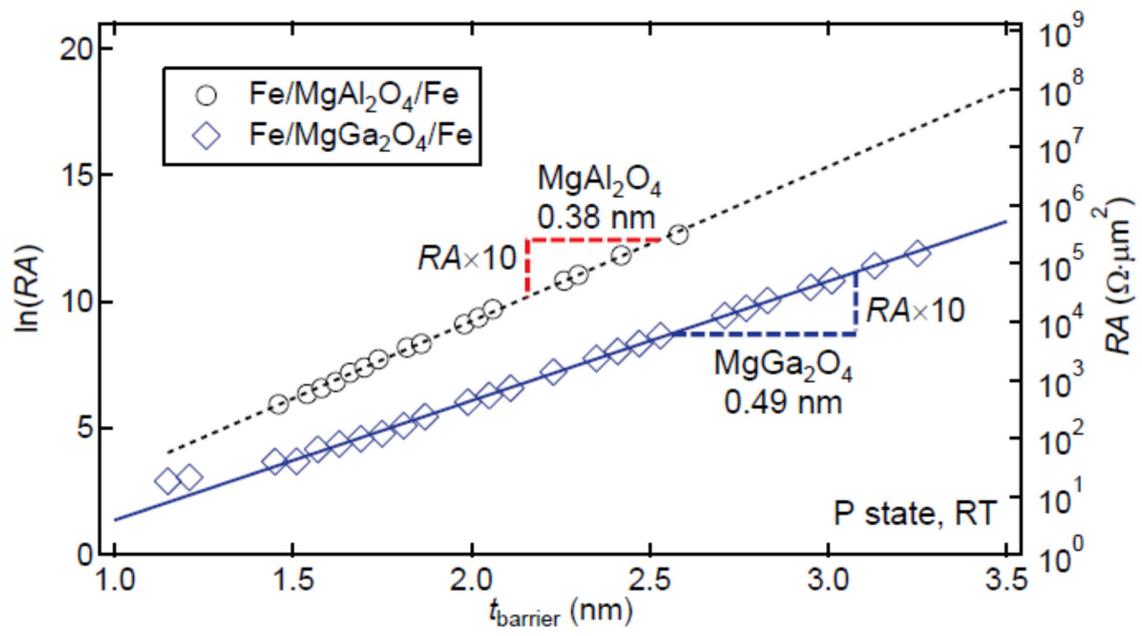

FIG. 3.

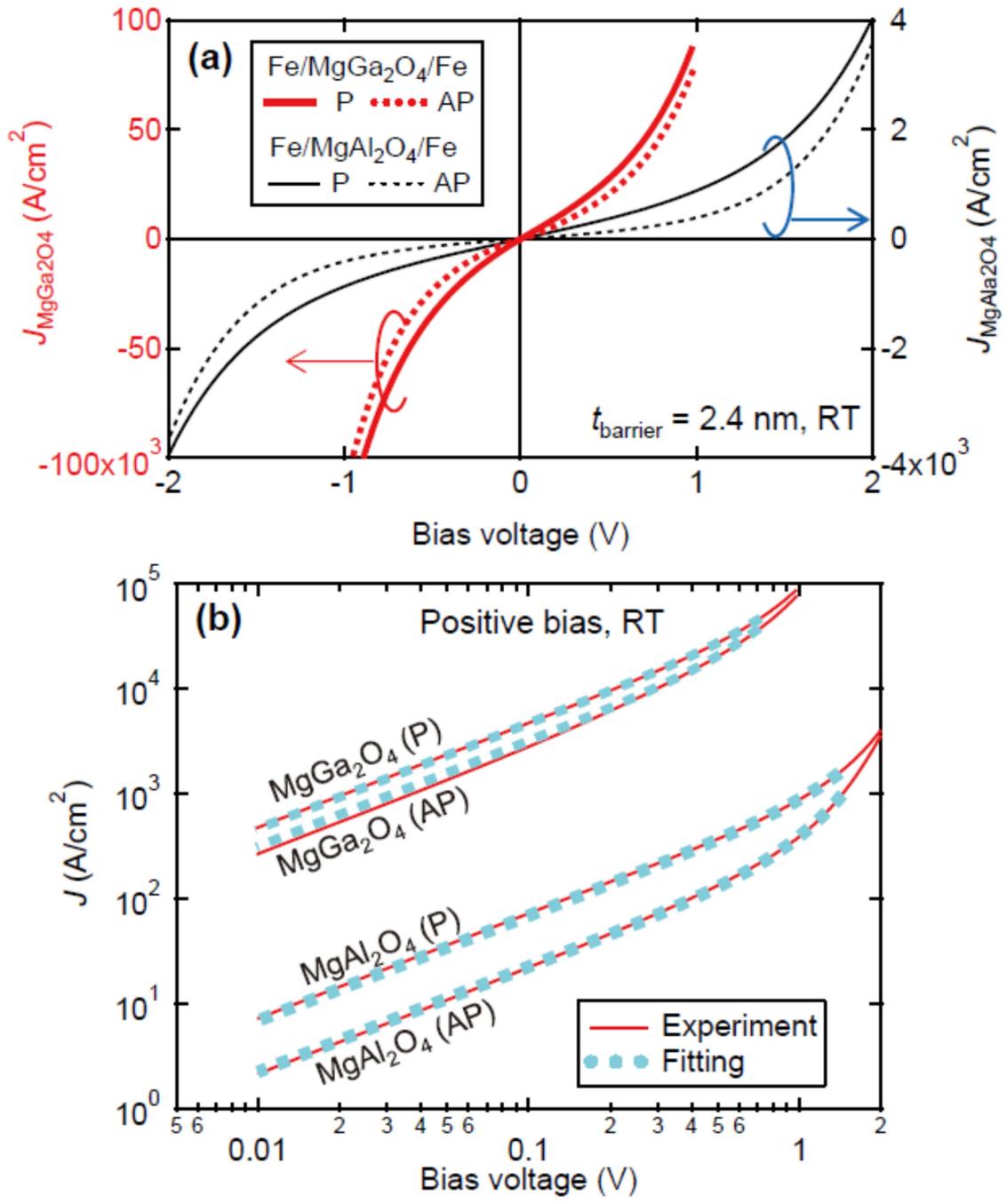

FIG. 4.